\documentstyle[11pt,newpasp,twoside,epsf]{article} 
\def\edcomment#1{\iffalse\marginpar{\raggedright\sl#1\/}\else\relax\fi} 
\reversemarginpar 

\begin{document} 
\title{FUSE Observations of the Post-AGB Star ZNG~1 in the Globular Cluster M5 (NGC 5904)}

\author{W. Van Dyke Dixon} 
\affil{Department of Physics \& Astronomy, The Johns Hopkins University,
3400 N. Charles Street, Baltimore, MD 21218, USA}

\author{Thomas M. Brown} 
\affil{Space Telescope Science Institute, 3700 San Martin Drive,
Baltimore, MD 21218, USA} 

\author{Wayne B. Landsman} 
\affil{Science Systems and Applications, Inc., Code 681, NASA Goddard
Space Flight Center, Greenbelt, MD 20771, USA}

\begin{abstract} 
We have observed the hot post-AGB star ZNG~1 in the globular cluster M5
with the {\it Far Ultraviolet Spectroscopic Explorer (FUSE).} From
the resulting spectrum, we derive an effective temperature $T_{\rm eff}
\sim 45,000$ K, a rotational velocity $v_{\rm rot} \sim 100$ km
s$^{-1}$, carbon and nitrogen abundances approximately ten times solar,
a wind velocity $v_{\infty} \sim 1000$ km s$^{-1}$, and evidence for an
expanding shell of material around the star.  The carbon and nitrogen
enhancements suggest dredge-up of nuclear-processed material on the
AGB.  The high rotational velocity may reflect a previous merger with a
binary companion.

\end{abstract}

\noindent
As part of a project to study mixing and mass loss in the atmospheres
of asymptotic-giant-branch (AGB) stars, we have observed the post-AGB
star ZNG~1 ({Zinn}, {Newell}, \& {Gibson} 1972) in M5 (NGC 5904) with
the {\it Far Ultraviolet Spectroscopic Explorer (FUSE).}  The resulting
spectrum, presented in Fig.~1, suggests that the star has experienced
both processes -- and that the mass loss continues.

The upper two panels show narrow absorption features of H {\small I}
and C {\small II} blueshifted by $\sim$ 190 km s$^{-1}$ relative to the
star.  C {\small III} absorption at the same velocity is also present,
but is not shown here.  These features may represent an
intermediate-velocity cloud or a shell of material previously ejected
by the star.  Broad P-Cygni profiles can be seen in both components of
the O {\small VI} doublet (middle panel), each punctuated by a discrete
absorption component (DAC) blueshifted by 920 km s$^{-1}$ relative
to the star.  The redder O {\small VI} doublet is interstellar.
Photospheric features of nitrogen and carbon can be seen in the bottom
panel.  Their redshift (60 km s$^{-1}$ relative to the ISM) is
consistent with that of the cluster (Harris 1996).  The star's iron
abundance is close to the cluster mean ([Fe/H] = $-1.29$; Harris 1986),
but its carbon and nitrogen abundances are ten times solar, suggesting
that products of helium burning were mixed to the surface while the
star was on the AGB.  The star's photospheric features are

\noindent
\parbox{5.25in}{\epsfxsize=5.25in \epsfbox{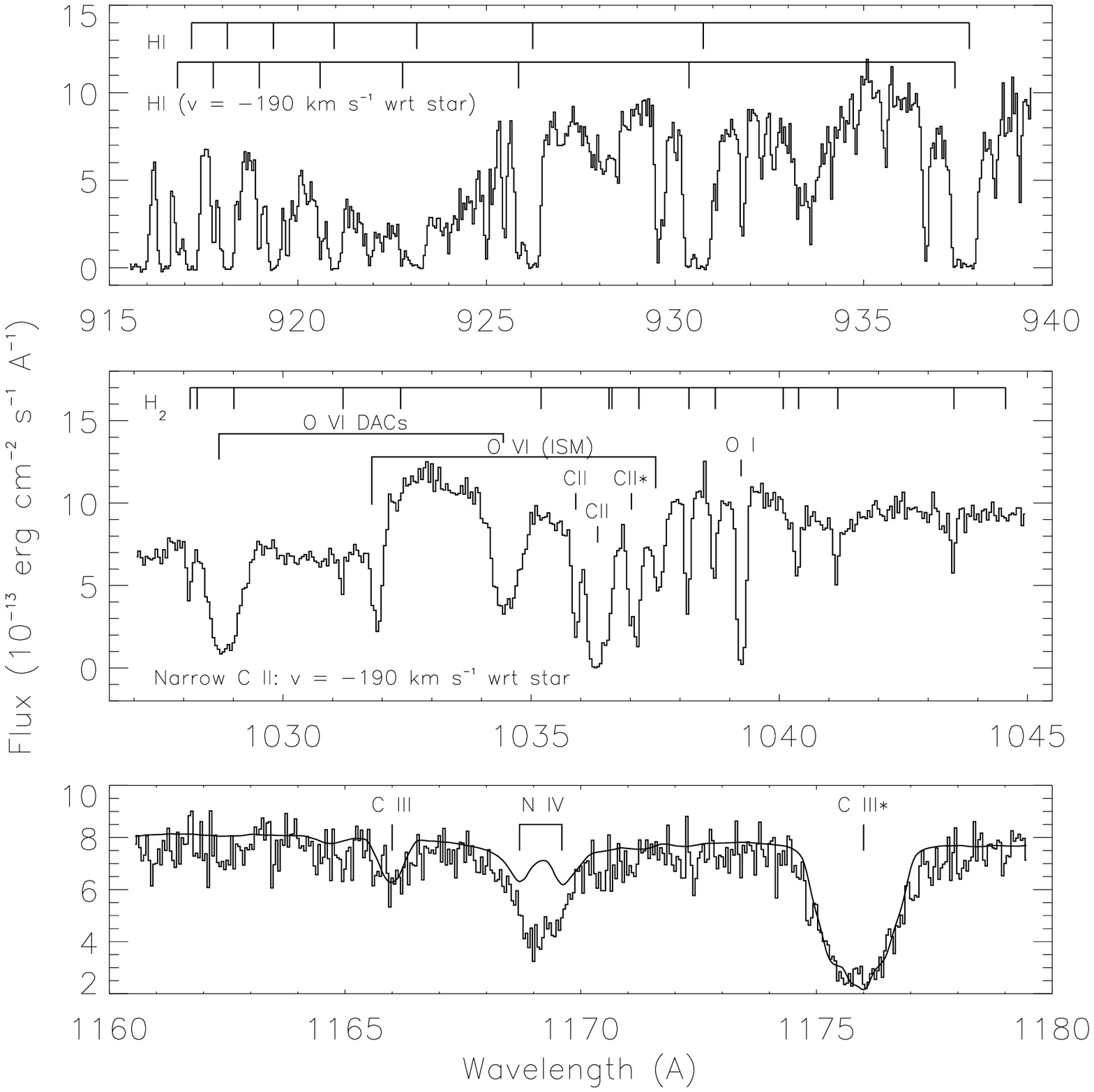}}

\vskip 0.1in
\noindent
\parbox{5.25in}{
Figure 1.  Far-UV spectrum of ZNG~1 in M5 obtained with {\it FUSE.} The
top and middle panels present evidence for a high-velocity stellar wind
and expanding shell.  In the bottom panel, a synthetic spectrum
(generated with the non-LTE code of Hubeny \& Lanz 1995) with $T_{\rm
eff} = 45,000$ K, $\log g = 4.5$, [Fe/H] = $-1.3$, and carbon and
nitrogen abundances ten times solar is plotted over the data.  A higher
nitrogen abundance is inconsistent with other features not shown here.
}

\bigskip
\noindent
well fit by a rotational velocity of $\sim$ 100 km s$^{-1}$, remarkably
fast for a post-AGB star.  The star may have been spun up in a merger
with a binary companion.


\begin{references}


\reference {Harris}, W.~E. 1996, \aj, 112, 1487


\reference {Hubeny}, I. \& {Lanz}, T. 1995, \apj, 439, 875

\reference {Zinn}, R.~J., {Newell}, E.~B., \& {Gibson}, J.~B. 1972, \aap, 18, 390

\end{references}
\end{document}